\documentclass{article}
\usepackage{spconf,amsmath,graphicx}
\usepackage{amssymb}
\usepackage{xcolor}
\usepackage{hyperref}
\hypersetup{
    colorlinks=true,
    linkcolor=blue,
    filecolor=magenta,      
    urlcolor=cyan,
}
\usepackage{booktabs}
\usepackage{caption}
\usepackage{subcaption}
\usepackage{url}
\newcommand{\accessedDate}{Oct.~2020}
\newcommand{\urlfootnote}[1]{\footnote{\url{#1} Accessed \accessedDate}}

\title{Crowdsourcing approach for subjective evaluation of echo impairment}
\name{$^1$Ross Cutler, $^2$Babak Naderi, $^1$Markus Loide, $^1$Sten Sootla, $^1$Ando Saabas}
\address{$^1$Microsoft Corporation, $^2$Technische Universit\"at Berlin}

\begin{document}
%
\maketitle
\begin{abstract}
The quality of acoustic echo cancellers (AECs) in real-time communication systems is typically evaluated using objective metrics like ERLE \cite{G168} and PESQ \cite{p862}, and less commonly with lab-based subjective tests like ITU-T Rec. P.831 \cite{ITU-P831}. We will show that these objective measures are not well correlated to subjective measures. We then introduce an open-source crowdsourcing approach for subjective evaluation of echo impairment which can be used to evaluate the performance of AECs. We provide a study that shows this tool is accurate and highly reproducible. This new tool has been recently used in the ICASSP 2021 AEC Challenge \cite{sridhar2020icassp} which made the challenge possible to do quickly and cost effectively. 
\end{abstract}
\section{Introduction}

A common speech enhancement component in communication devices is the acoustic echo canceller (AEC), which attempts to remove the loudspeaker signal from the microphone signal. AECs are required not only for speakerphones, but also for headsets, handsets, and networks. The degradation of call quality due to acoustic echoes is one of the major sources of poor speech quality ratings in voice and video calls. As AECs are such a fundamental technology to telephony systems, there are many metrics to try to measure the performance of an AEC.

Most AEC publications (e.g., \cite{zhang2018deep, fazel2019deep, guerin2003nonlinear}) use objective measures such as echo return loss enhancement (ERLE) \cite{G168} and perceptual evaluation of speech quality (PESQ) \cite{p862}. ERLE is defined as: 

\begin{equation}
ERLE \approx 10\log_{10} \frac{\mathbb{E}[y^2(n)]}{\mathbb{E}[e^2(n)]} 
\end{equation}

\noindent where $y(n)$ is the microphone signal,  and $e(n)$ is the residual echo after cancellation. ERLE is only appropriate when measured in a quiet room with no background noise and only for single talk scenarios (not double talk). PESQ has also been shown to not have a high correlation to subjective speech quality in the presence of background noise \cite{Avila2019}.
Traditionally, the Quality of Experience (QoE) is measures by conducting subjective test in the laboratory following the standardized test methods \cite{naderi2020towards}.  
The ITU-T Rec. P.831 \cite{ITU-P831} provides guidelines on how to conduct subjective tests for network echo cancellers in the laboratory. 
ITU-T Rec. P.832 \cite{itut_p832} focuses on the hands-free terminals and cover a broader range of degradations. However due to the high costs of subjective experiments in the laboratory usage of objective metrics are increased.

There are many standards for AEC performance. IEEE 1329 \cite{ieee1329} defines metrics like terminal coupling loss for single talk (TCLwst) and double talk (TCLwdt), which are measured in anechoic chambers. 
TIA 920 \cite{tia920} uses many of these metrics but defines required criteria. 
ITU-T Rec. G.122 \cite{g122} defines AEC stability metrics, and ITU-T Rec. G.131 \cite{ITU-G131} provides a useful relationship of acceptable Talker Echo Loudness Rating versus one way delay time.
ITU-T Rec. G.168 \cite{G168} provides a comprehensive set of AEC metrics and criteria. However, it is not clear how to combine these dozens of metrics to a single metric, or how well these metrics correlate to subjective quality. 

Our goal is to develop a minimal set of end-to-end subjective tests to characterize an AEC for developing deep AEC's and conducting an AEC challenge. The tests need to be able to stack rank the performance of AEC's accurately and reliably, and be cost effective. 

In Section \ref{sec:implementation} we describe the implementation of our subjective methods for evaluating echo impairment adapted from the ITU-T Recommendations P.831, P.832, and P.808. In Section \ref{sec:comparison} we show ERLE and PESQ don't correlate well to subjective scores. In Section \ref{sec:accuracy} we show the crowdsourcing test has good accuracy compared to expert raters. In Section \ref{sec:reproducibility} we show our implementation is highly reproducible. In Section \ref{sec:challenge} we show how our subjective metrics were used in a challenge to determine the best performing models. 

\section{Implementation}
\label{sec:implementation}
We have extended the open source P.808 Toolkit \cite{naderi2020open} with methods for evaluating the echo impairments in subjective tests. We followed the \textit{Third-party Listening Test B} from ITU-T Rec. P.831 \cite{ITU-P831}, and ITU-T Rec. P.832 \cite{itut_p832} and adapted them to our use case as well as for the crowdsourcing approach based on the ITU-T Rec. P.808 \cite{ITU-P808} guidance. Our implementation is available here\urlfootnote{https://github.com/microsoft/P.808}.

A third-party listening test differs from the typical listening-only tests (according to the ITU-T Rec. P.800) in the way that listeners hear the recordings from the \textit{center} of the connection rather in the former one in which the listener is positioned at one end of the connection \cite{ITU-P831}. Thus, the speech material should be recorded by having this concept in mind.
During the test session,  we used different combinations of single- and multi-scale Absolute Category Rating (ACR) ratings depending on the speech sample under evaluation. We distinguished between single talk and double talk scenarios.
For the near end single talk, we asked for the overall quality, but for far end single talk and double talk scenarios we asked for echo annoyance and impairments from other degradations in two separate questions (see Table \ref{table:survey}). Both impairments were rated on the degradation category scale (from 1:\textit{Very annoying}, to 5: \textit{Imperceptible}). The impairments scales leads to a Degradation Mean Opinion Scores (DMOS). We asked participants to rate the other degradations as the AEC, like any other speech enhancement technique, can also add new speech transmission impairments \cite{ITU-P831}. 

\begin{table}[t]
\centering
    \begin{tabular}{|p{0.2\linewidth} | p{0.7\linewidth}|}
        \hline
        Scenario & Question \\ \hline
        Single talk & 1. How would you rate the degradation from acoustic echo in this speech sample? \\ 
         & 2. How would you judge other degradations (noise, distortions, etc.) of this speech sample? \\ \hline
         Double talk &  1. How would you judge the degradation from the echo of Person 1's voice? \\
         & 2. How would you judge degradations (missing audio, distortions, cut-outs) of Person 2's voice? \\ \hline
    \end{tabular}
    \caption{Test items used in different scenarios of subjective tests.}
    \label{table:survey}
\end{table}

\subsection{Speech material}
\begin{figure}[t]
	\includegraphics[width=1\columnwidth]{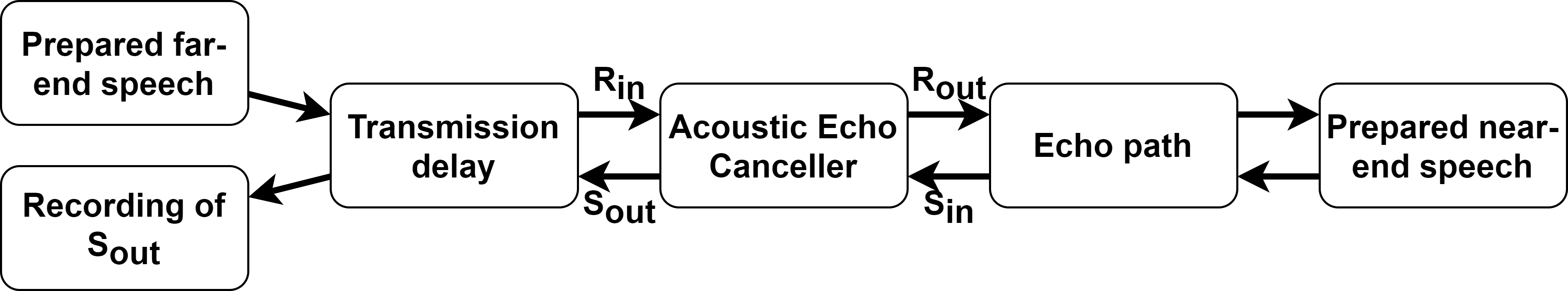}
	\caption{Echo canceller test set-up for Third Party Listening Test B according to the ITU-T Rec.P.831 (after \cite{ITU-P831}). $S$ is send and $R$ is receive.}
	\label{fig:P831}
\end{figure}

\begin{figure}[t]
    \centering
	\includegraphics[width=0.7\columnwidth]{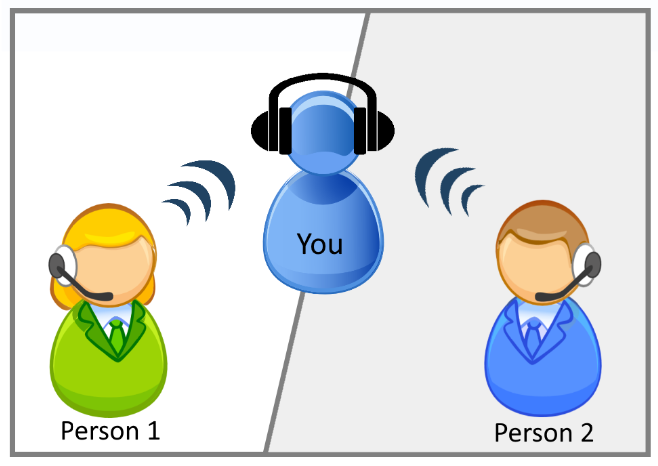}
	\caption{Double talk scenario in the Third Party Listening Test. The test participant (marked by "You") is positioned in the center of the communication.}
	\label{fig:double_talk}
\end{figure}

The setup illustrated in Figure \ref{fig:P831} should be used to process all speech samples with all of the AECs under the study. To simplify the rating process for crowdworkers, we distinguished between near end and far end single talk as well as the double talk scenarios and tried to simulate them for the test participants\footnote{As part of the dataset was collected through crowdsourcing using an artificial head measurement system for recording was not possible.}.
In the case of near end single talk we recorded the AEC output ($S_{out}$). For far end single talk, we added the output of the AEC ($S_{out}$) with a delay of 600ms to the loopback ($R_{in}$) signal, yielding $R_{in} + $ delayed $S_{out}$. For the listener, this simulates hearing the echo of their own speech (i.e., $R_{in}$ as an acoustic sidetone). For double talk the process is similar, but due to there being more speakers, simply adding the delayed AEC output ($S_{out}$) would cause confusion for the test participants. To mitigate this issue, the signals are played in stereo instead, with the loopback signal ($R_{in}$) played in one ear (i.e., acoustic sidetone) and the delayed output of the AEC ($S_{out}$) played in the other. Figure \ref{fig:double_talk} was used to illustrate the double talk scenario to crowdworkers.

\subsection{Crowdsourcing test}
Our extension followed the main structure of the P.808 Toolkit. The toolkit is highly automated to avoid operational errors. It contains scripts for creating the crowdsourcing test (generate the HTML file, trapping-stimuli set, input URLs, etc.) given the specific scenario and also a script for processing the submitted answers (i.e., data screening and aggregating the reliable ratings). The speech material should be prepared separately following the above-mentioned guidance.

The crowdsourcing test includes several sections from the participant's perspective. The \textit{Qualification} section includes a hearing ability of test participants using a digit-triplet test. In the \textit{Setup} section, usage of both ear-pods and suitability of the participant's environment are evaluated using the modified just-noticeable difference in quality test method \cite{naderi2020env}. In the \textit{Training} section, the test participant is introduced to and use the single- or multi-scale rating procedure for rating a predefined set of speech samples. The samples in the training set should be carefully selected and cover the entire range of the scales. A simplified instruction about the third-party listening test is also provided in this section. The last section is the \textit{Ratings} section in which participants listen to a set of stimuli and rate them on the given scales.
As the crowdsourcing task should be kept short, it is recommended to only use about ten speech samples in the rating section. Participants can perform multiple tasks. In that case, the \textit{Qualification} section only appears in the first task. The setup and training sections appear periodically (the default setting is every 30 and 60 minutes, respectively) when the worker passed them successfully.  

In case multiple scales are used, participants should listen to the speech sample again, attending only to the specified aspect of that scale before voting. The presentation order of scales is also randomized in each task to avoid any order effect.

In every crowdsourcing task, it is recommended to include gold standard and trapping questions \cite{naderi2018motivation, naderi2015effect}. The trapping stimuli is an obvious quality control mechanism \cite{naderi2015effect} which asks the participant to select a specific response to show their attention. The trapping question asks the participant to either select a specific category's label (in case of one-scale rating) or specific score (multi-scale rating).
We kept the structure of the P.808 Toolkit the same. Further details about the P.808 Toolkit can be found in \cite{naderi2020open} and details on validation of the ITU-T Rec. P.808 in \cite{naderi2020dtowards}. 

\section{Comparison to objective metrics}
\label{sec:comparison}
Next, we measure the correlation between the subjective metrics as evaluated with our proposed methodology and commonly used objective metrics from the literature, ERLE and PESQ.

Specifically, we took the 400 far-end single talk outputs from each of the top 16 AEC models the ICASSP 2021 Acoustic Echo Cancellation Challenge \cite{sridhar2020icassp} (excluding the last-ranking outlying model to reduce the spread -- see Table \ref{fig:challenge}), and from 3 internal models. In total this was 7,600 enhanced speech signals. We obtained their subjective DMOS ratings using our tool and then calculated their corresponding objective metrics. Table \ref{table:1} gives the correlations between those measures across the 7,600 far-end single talk datapoints.

\begin{table}[h!]
\centering
    \begin{tabular}{| c | c | c |}
        \hline
        & PCC & SRCC \\ \hline
        ERLE & 0.494 & 0.510 \\ \hline 
        PESQ & 0.637 & 0.533 \\ \hline
    \end{tabular}
    \caption{PCC and SRCC between subjective DMOS collected by our toolkit and objective measures for far end single talk.}
    \label{table:1}
\end{table}

The correlations between the subjective and objective metrics are too weak to stack-rank AEC models to determine which are the best performers. 
\section{Accuracy study}
\label{sec:accuracy}
To assess the accuracy of the subjective test we conducted a study comparing the results of expert raters and crowdsourced raters. The expert raters were $N=5$ Microsoft engineers working on audio processing. The study used the near end single talk test set from \cite{sridhar2020icassp}. The results in Table \ref{table:accuracy} show a significant correlation improvement by using a two question scale (given in Table \ref{table:survey}) which separates echo and other distortions, compared to an echo only scale (just question 1 in Table \ref{table:survey}). As a result we use the two question scale. Table \ref{table:accuracy} also shows the correlation for different clips i.e. noisy, clean, and all clips; the best results are with clean clips. Overall the study shows that the crowdsourced raters have an acceptable correlation (PCC=0.85) with the expert raters when using the two question scale. 

\begin{table}[t]
\centering
    \begin{tabular}{| c | c | c | c |}
        \hline
        Scenario & Noisy & Clean & All \\ \hline
        Single question survey & 0.68 & 0.83 & 0.70 \\ \hline
        Two question survey & 0.84 & 0.89 & 0.85 \\ \hline
    \end{tabular}
    \caption{Correlation between expert and crowdsourced raters}
    \label{table:accuracy}
\end{table}

\section{Reproducibility study}
\label{sec:reproducibility}
\begin{figure}[t]
	\includegraphics[width=1\columnwidth]{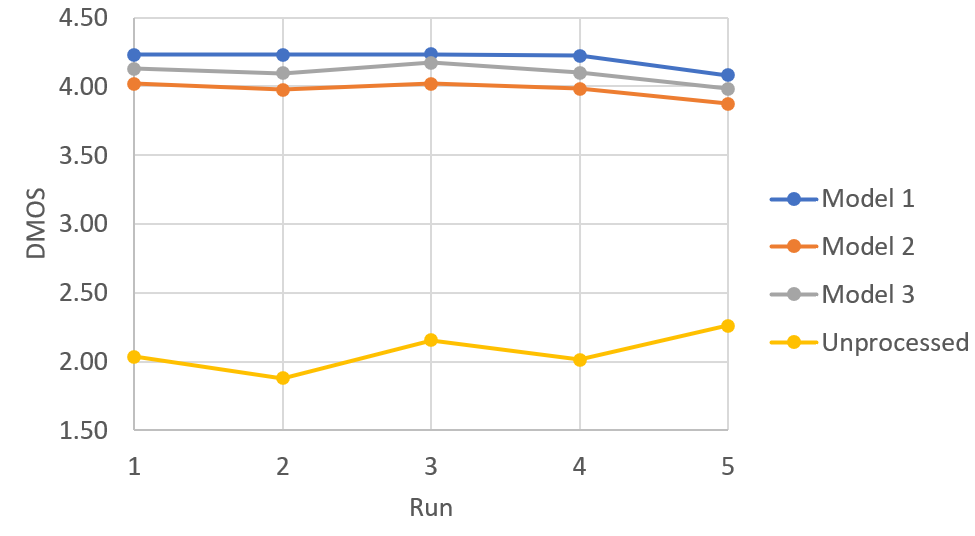}
	\caption{Far end single talk echo DMOS of five consecutive test runs.}
	\label{fig:st_echo}
\end{figure}

\begin{figure}[t]
	\includegraphics[width=1\columnwidth]{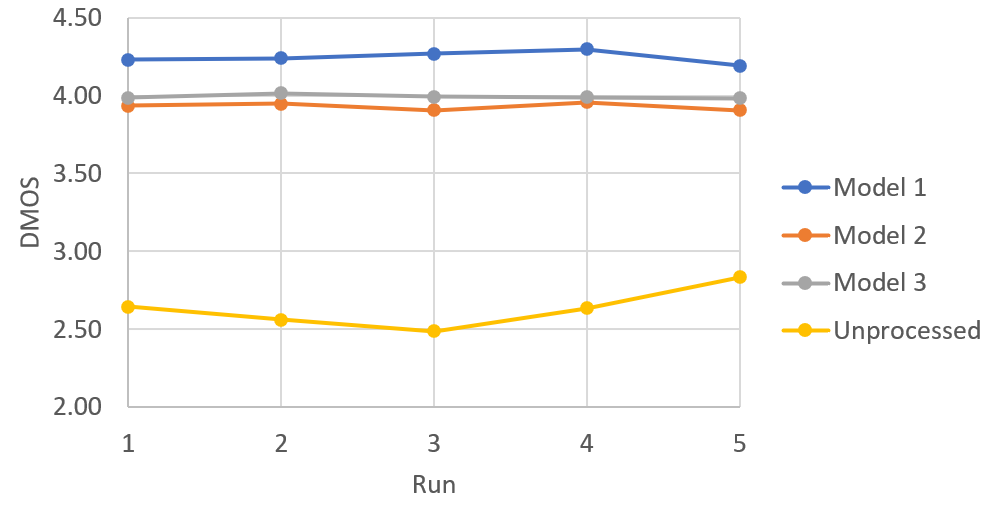}
	\caption{Double talk echo DMOS of five consecutive test runs.}
	\label{fig:dt_echo}
\end{figure}

\begin{figure}[t]
	\includegraphics[width=1\columnwidth]{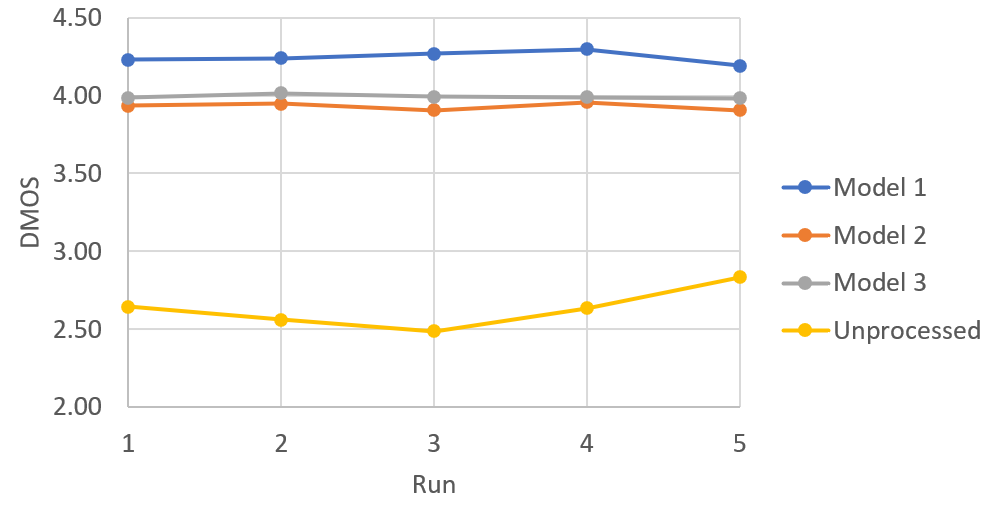}
	\caption{Double talk other degradation DMOS of five consecutive test runs.}
	\label{fig:dt_other}
\end{figure}

To assess the reproducibility of subjective measurements using our toolkit, we conducted a study on the dataset used in the ICASSP 2021 AEC Challenge \cite{sridhar2020icassp}. We used 400 far end single talk and 400 double talk clips from the blind test set and ran separate tests with both, five times on five separate days, with far end single talk and double talk falling on different days. For the far end single talk runs, each run had unique raters, but for the double talk runs we allowed raters to participate again, to assess the impact of recurring versus distinct raters. In both cases, we saw similar results in terms of reproducibility (see below). For the far end single talk runs, an average of $N=86.2$ workers participated in each run. For the double talk runs, an average of $N=124.6$ workers participated per run. We used four test conditions, three internally developed AEC models, and an unprocessed reference condition. We targeted 10 votes per condition and received an average of 8.3, resulting in 95\% confidence intervals of $\sim$0.02. The results are shown in Figures \ref{fig:st_echo}, \ref{fig:dt_echo} and \ref{fig:dt_other}. 
Both the SRCC and the PCC were 1.00 for all test runs, showing excellent reproducibility (Table \ref{table:pcc_scc_repro}). 

\begin{table}[t]
\centering
    \begin{tabular}{| c | c | c |}
        \hline
        Scenario & PCC & SRCC \\ \hline
        FE ST Echo & 1.00 & 1.00 \\ \hline
        DT Echo & 1.00 & 1.00 \\ \hline
        DT Other & 1.00 & 1.00 \\ \hline
    \end{tabular}
    \caption{PCC and SRCC for FE ST echo, DT Echo and other degradation scenarios between five runs.}
    \label{table:pcc_scc_repro}
\end{table}

\section{AEC Challenge study}
\label{sec:challenge}
The 2021 ICASSP AEC Challenge \cite{sridhar2020icassp} included 17 participants who processed 1,000 scenarios on a blind test set. We also included a baseline model in the evaluation, for a total of 18 models. The results are shown in Table \ref{fig:challenge}. The columns are:

\begin{itemize}\itemsep0pt
    \item ST NE MOS: P.808 overall quality MOS of near end single talk scenario
    \item ST FE Echo DMOS: Echo annoyance DMOS for far end single talk
    \item DT Echo DMOS: Echo annoyance DMOS for double talk scenario
    \item DT Other DMOS: Other impairments DMOS of double talk scenario
    \item Overall: Mean of MOS and DMOS metrics
\end{itemize}

The evaluation was broken into 3 subjective tests using our Toolkit, one P.808 ACR test with a near end single talk scenario, one single-scale echo annoyance test for the far end single talk, and one multi-scale echo annoyance test for the double talk scenarios. We targeted 10 votes per clip
and received an average of 8.5. For each of the tests we got N=137, N=238, N=236 raters. 
For the final result (Overall) the 95\% confidence interval is 0.02. The single talk ACR test was completed in less than 1 day, while the double talk scenario took 3 days. The crowdworkers were compensated by \$0.5 for the single talk test (with 10 clips), and \$0.9 for the double talk test (with 12 clips).

\begin{table}[t]
	\includegraphics[width=1\columnwidth]{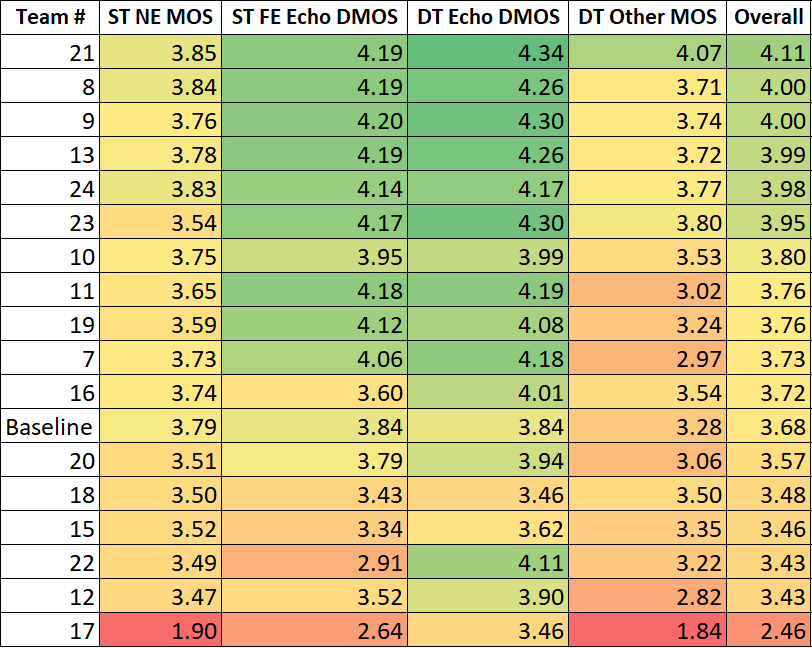}
	\caption{AEC Challenge results.}
	\label{fig:challenge}
\end{table}

\section{Discussion and Conclusion}
\label{discussion}
We have provided a cost effective, accurate, and reproducible set of subjective tests to measure end-to-end performance of AECs, which enables not just AEC challenges but also the development of deep AECs that optimize these metrics. Future research includes investigating adding additional questions to measure loudness variation and an overall quality of experience that better aggregates the four metrics we currently measure. 

\bibliographystyle{IEEEbib}
\bibliography{strings,refs}

\end{document}